\documentclass[conference,10pt]{IEEEtran}
\usepackage[left=1.54cm,right=1.54cm,top=1.8cm,bottom=4.21cm]{geometry}
\usepackage{cite}
\usepackage{amsmath,amssymb,amsfonts,bbm}
\usepackage{graphicx}
\usepackage{textcomp}
\usepackage[dvipsnames]{xcolor}
\usepackage{multirow}
\usepackage[nolist]{acronym}
\usepackage{subcaption}
\usepackage{caption}
\usepackage{tikz}
\usepackage{tkz-tab}
\usetikzlibrary{automata,arrows,positioning,calc}
\usetikzlibrary{shapes,snakes}
\usetikzlibrary{arrows}
\usepackage{dsfont}
\usepackage{soul}
\usepackage{subfiles}
\usepackage{comment}

\usepackage{algorithm}
\usepackage{algpseudocode}
\makeatletter 
\algrenewcommand\ALG@beginalgorithmic{\footnotesize}
\makeatother
\usepackage{amsmath}
\usepackage{amssymb}
\usepackage{amsthm}

\usepackage{booktabs}

\newcommand{\figref}[1]{\figurename~\ref{#1}}

\colorlet{pink}{red!40}
\colorlet{blue}{cyan!50}

\usepackage{tikz}
\usetikzlibrary{fit,calc}

\colorlet{pink}{red!40}
\colorlet{blue}{cyan!50}

\begin{document}

\bstctlcite{IEEEexample:BSTcontrol}

\title{Continuous Multi-Link Operation: A Contention-Free Mechanism for the Unlicensed Spectrum}

\author{
\IEEEauthorblockN{Gianluca Fontanesi$^{\star}$, Francesc Wilhelmi$^{\star}$, and Lorenzo Galati-Giordano$^{\star}$ \vspace{0.1cm}
}
\IEEEauthorblockA{$^{\star}$\emph{Radio Systems Research, Nokia Bell Labs, Stuttgart, Germany}}
}

\maketitle
\thispagestyle{empty} 
\pagestyle{plain}

\begin{abstract}
This paper proposes a novel mechanism to enforce contention-free channel access in the unlicensed spectrum, as opposed to the traditional contention-based approach. To achieve this objective, we build on the Wi-Fi~7 multi-link operation (MLO) and define the means whereby independent channel access attempts are performed in all the addressable links to ensure one available channel/link is ready for transmission at all times, such that a sequence of continuous acquired channels can be maintained. We call this method continuous multi-link operation (ConMLO). In this work, we aim to verify the applicability of ConMLO, its ability to retain spectrum resources for a given duration of time, and its fairness with respect existing approaches, namely legacy single-link operation (SLO) and MLO. To this end, we use realistic data traffic measurements acquired in a crowded football stadium as an exemplary case of challenging spectrum occupation. Our results show that the proposed ConMLO can effectively guarantee continuous channel acquisition under different occupancy scenarios without compromising fairness of channel access compared to existing legacy modes. 
\end{abstract}


\begin{acronym}
\acro{AIFS}{arbitration interframe space}
\acro{AP}{access point}
\acro{AR}{augmented reality}
\acro{CA}{collision avoidance}
\acro{CSMA}{Carrier Sense Multiple Access}
\acro{Cont-STR}{continuous multi-link operation}
\acro{Cont-TX}{continuous transmission}
\acro{Cont-MLChA}{continuous multi-link channel access}
\acro{ConMLO}{continuous multi-link operation}
\acro{CW}{contention window}
\acro{DCF}{distributed coordination function}
\acro{ED}{energy detection}
\acro{EMLMR}{enhanced multi-link multi-radio}
\acro{EMLSR}{enhanced multi-link single-radio}
\acro{LBT}{listen-before-talk}
\acro{MAC}{medium access control}
\acro{MLD}{multi-link device}
\acro{MLO}{multi-link operation}
\acro{MLO-STR}{multi-link operation with simultaneous transmit and receive}
\acro{MLO-NSTR}{multi-link operation with non-simultaneous transmit and receive}
\acro{ML}{machine learning}
\acro{NSTR}{non-simultaneous transmit and receive}
\acro{RSSI}{received signal strength indicator}
\acro{RTA}{real-time application}
\acro{SLO}{single-link operation}
\acro{STR}{simultaneous transmit and receive}
\acro{STA}{station}
\acro{TXOP}{transmission opportunity}
\acro{UHR}{ultra-high-reliability}
\acro{URLLC}{ultra-reliable low-latency communication}
\acro{VR}{virtual reality}
\acro{WACA}{Wi-Fi all-channel analyzer}
\acro{MAPC}{multi-access point coordination}
\end{acronym}

\section{Introduction}
\label{sec:introduction}

Next-generation communications systems will have to support the challenging requirements in terms of throughput, low-latency, and reliability of emerging applications such as \ac{VR}, \ac{AR}, or Metaverse. To such end, the release of new frequency bands has been identified as key for supporting these bandwidth-hungry applications. In this regard, the license-exempt spectrum (alternatively, the unlicensed spectrum) is foreseen to play an important role, thus granting communications systems plenty of bands for supporting novel use cases. 

The unlicensed approach, however, enforces specific rules to ensure fairness in the way different technologies and devices are sharing the spectrum resources, which in turn are detrimental to reliability. Listen-before-talk (LBT) is typically adopted for devices operating in the unlicensed spectrum and willing to gain channel access. Through \ac{LBT}, a device, before transmitting, must first listen to the medium to ensure that no other devices are currently occupying the spectrum, so that collisions are avoided, and then occupy the frequency resources for the duration of a \ac{TXOP},  typically a few milliseconds. The main drawback of \ac{LBT}-based methods is precisely in its intrinsic contention, which may lead to unpredictable transmission delays and unreliability as the number of operating devices in the same coverage area increases. 

With around 20 billion devices actively in use in 2023, Wi-Fi is one of the main technologies operating in the unlicensed spectrum. Since its first release in 1999, Wi-Fi has consistently evolved throughout its generations towards improving its classical target, i.e. peak data rate, from 1 Mbps in the first release of the initial IEEE 802.11 standard to 36 Gbps in the latest IEEE 802.11be amendment (Wi-Fi 7). However, beyond peak data rate, the next generation of Wi-Fi, the IEEE 802.11bn (Wi-Fi 8)~\cite{giordano2023will}, is set to target, for the first time, \ac{UHR} in the communications. To meet the reliability and latency requirements of emerging applications, new advanced functionalities are currently under study, like \ac{MAPC} (e.g., coordinated spatial reuse~\cite{wilhelmi2023throughput}), or the extension of \ac{MLO}, firstly introduced in Wi-Fi 7, targeting a more dynamic and efficient use of the frequency channels \cite{carrascosa2023understanding}. The latter feature allow a \ac{STA} to establish connection over all the available links/channels through a single association.

In this paper, we aim to introduce a novel approach referred to as \ac{ConMLO} for bypassing the main source of unreliability in the wireless unlicensed, namely contention, and maintaining constant channel acquisition through a set of links.
This can be seen as an extension of \ac{MLO} in the direction of acquiring, following the \ac{LBT} rules, consecutive \ac{TXOP}s spread over different links, and taking advantage of the degree of freedom introduced by \ac{MLO}. 
In order to assess the feasibility of our proposed approach, we tested its behaviour by emulating its possible deployment  in a realistic environment by using the openly available measurement campaign of the \ac{WACA} dataset~\cite{barrachina2021wi}. 
The proposed \ac{ConMLO} achieves high airtime and continuous channel acquisition under realistic channel occupancy, efficiently overcoming transmission interruptions in unlicensed spectrum systems without compromising fairness of channel access  compared to existing legacy modes.

\section{Related Work}
\label{sec:related_work}
\thispagestyle{empty} 
In line with the objective to improve reliability and latency of communication in the unlicensed spectrum, numerous approaches are available in the literature. 

Research debates have begun to develop novel channel access rules to be adopted in future available unlicensed spectrum~\cite{fcc2023fact}. However, although starting from a clean state may foster the introduction of completely new paradigms beyond the traditional \ac{LBT} approach, this would not benefit deployments operating in current unlicensed bands. 

Driven by this observation, proposals to improve unlicensed channel access by tweaking \ac{LBT} consider adopting a dynamically adjusted \ac{ED} thresholds~\cite{afaqui2015evaluation}, also via centralized or distributed \ac{ML} algorithms \cite{zhou2024federated}. A recent work~\cite{wilhelmi2024conpa} proposes 
to remove channel contention and its associated unreliability by adapting the transmission power based on the interference measured during \ac{ED}. In this way, higher transmit power is allowed when lower interference is measured. Although this approach may result simple and effective, its applicability is limited to new spectrum rules.

Traffic prediction can also help in creating solutions that shorten the channel access delay for current \ac{LBT}-based mechanisms. In this case, if a device is capable of anticipating the arrival of a certain data traffic, it can prepare for accessing the channel in advance to meet latency goals. An example of predictive channel access is proposed in \cite{chemrov2022smart}, which considers accessing the channel in advance before the arrival of \ac{RTA} data thus to reserve the channel for immediate transmission of delay-sensitive traffic. The disadvantage of this type of approaches is that they increase the complexity and cost of the devices.

Finally, when it comes to the \ac{MLO} functionality on top of which we develop our method, its performance has been extensively studied in \cite{carrascosa2023understanding}. \ac{MLO} is shown to be an appealing solution to reduce the transmission delay by attempting channel access simultaneously in different links and decide to transmit, e.g., in the first accessible one. However, contrary with the overarching goal of our proposed solution, 
there is no guarantee that consecutive \ac{TXOP} can be enforced and enable contention-free transmissions.

\section{Continuous Multi-Link Operation (ConMLO)}\label{ContxIdea}
\thispagestyle{empty} 

\begin{figure}[t!]
    \centering
    \includegraphics[width=.9\linewidth]{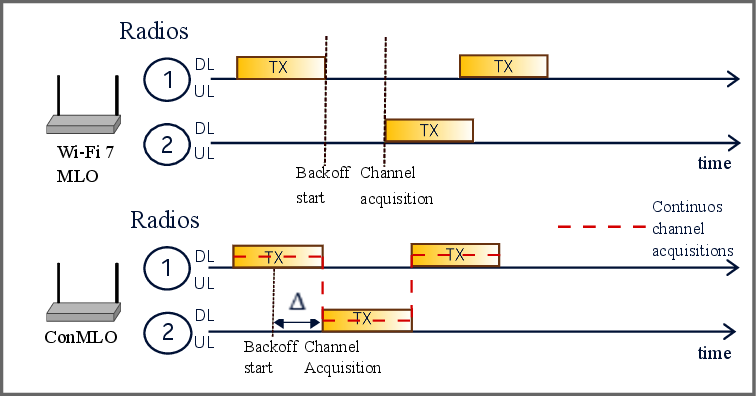}
    \caption{The operation of Wi-Fi 7 \ac{MLO} and the proposed \ac{ConMLO}. While Wi-Fi 7 \ac{MLO} needs to restart the channel access after finalizing a transmission, \ac{ConMLO} provides continuity by anticipating the channel access by $\Delta$ in other links.}    
    \label{fig:CTX_operation}
\end{figure}

To address the unpredictable delays and lack of reliability due to channel access contention, we propose \ac{ConMLO}, a mechanism that aims to provide continuous, uninterrupted channel access by performing opportunistic frequency hopping across different links/channels of the available spectrum. \ac{ConMLO} is based on the existing Wi-Fi~7 \ac{MLO} feature but, unlike existing \ac{MLO} methods~\cite{carrascosa2023understanding}, it coordinates channel access contention across the different links to enforce continuity of channel acquisition. \figref{fig:CTX_operation} outlines the operation of both Wi-Fi 7 MLO (above) and \ac{ConMLO} (below) for a \ac{MLD} operating in two different links. In the following, we introduce the basics of the \ac{MLO} framework and describe the proposed \ac{ConMLO} mechanism designed on top of it.

\subsection{\ac{MLO} basics}

\ac{MLO} is the most innovative feature that Wi-Fi 7 has brought along. Based on this feature, Wi-Fi devices can operate multiple channels/links jointly, so that the different radios available in a device (e.g., at 2.4~GHz, 5~GHz, and 6~GHz) can be leveraged to increase the performance of a given communication. In particular, Wi-Fi 7 defined a framework whereby multiple links can be used seamlessly through the same association, 
following different modes of operation, namely \ac{EMLSR} and \ac{EMLMR} \cite{carrascosa2023understanding}. In both cases, devices have the capability of simultaneously performing contention over different links, but only transmit using a single fully functional radio (i.e., \ac{EMLSR}) or multiple ones (i.e., \ac{EMLMR}). In this paper, for the sake of simplicity and results presentation but without loss of generality, we refer to EMLSR interchangeably as Wi-Fi 7 MLO.

\subsection{ConMLO formulation}

We assume a set $\mathcal{L} = 1,..., L$ of links in which a \ac{ConMLO} device can perform sensing simultaneously when attempting to initiate a transmission. For that, a backoff counter array $\mathbf{BO} \in \mathbb{Z}^l$ is maintained for each link $l\in\mathcal{L}$, where each $BO_l \in \mathbf{BO}$ is decreased independently by following the standard \ac{DCF} procedure. More specifically, for a specific link $l$, $BO_l$ can be decreased by one unit if link $l$ has been detected as idle ($\rho=0$), i.e., the power of interference sensed $P_\text{int}$ during that interval is below an \ac{ED} threshold $\phi$, during the last basic slot duration $\delta$. Otherwise, if the channel is sensed as busy ($P_{int} \geq \phi$), contention must be done, thus freezing the backoff counter to its current state. Before attempting to decrease a given backoff $BO_l$, an additional condition mandated by \ac{DCF} states that a preliminary channel sensing must be performed to ensure that channel $l$ has been idle during an \ac{AIFS} period, which includes an arbitrary number of slots $\delta$.

In \ac{ConMLO}, similar to Wi-Fi 7 MLO, a transmission of duration $T_\text{tx}$ is initiated over the link $l_w$ whose backoff counter expires first. In case of a tie, a random link among the winners is selected.  When a transmission is initiated, all backoff counters $BO_l \in \mathbf{BO}$ are reset to a random value between 0 and a \ac{CW} value. The main novelty that \ac{ConMLO} brings along with respect to Wi-Fi 7 MLO is that, to ensure continuity in the transmission, the backoff procedure is initiated before the end of the current transmission. More specifically, a shift $\Delta$ is defined to determine how far in advance the contention should be initiated  in the rest of the links $\mathcal{L}\setminus l_w$ before the end of the ongoing transmission in link $l_w$, so that a new transmission can be started immediately after the finalization of the current one. 

Algorithm~\ref{alg:dcf_11ax} shows the \ac{ConMLO} operation described above, for which a state machine $S=\{\text{TRANSMIT, SENSE, WAIT}\}$ is defined to describe the different behaviors of the device implementing \ac{ConMLO}.

\algnewcommand\algorithmicswitch{\textbf{switch}}
\algnewcommand\algorithmiccase{\textbf{case}}
\algnewcommand\Assert[1]{\State \algorithmicassert(#1)}%

\algdef{SE}[SWITCH]{Switch}{EndSwitch}[1]{\algorithmicswitch\ #1\ \algorithmicdo}{\algorithmicend\ \algorithmicswitch}%
\algdef{SE}[CASE]{Case}{EndCase}[1]{\algorithmiccase\ #1}{\algorithmicend\ \algorithmiccase}%
\algtext*{EndSwitch}%
\algtext*{EndCase}%

\algdef{SE}[SUBALG]{Indent}{EndIndent}{}{\algorithmicend\ }%
\algtext*{Indent}
\algtext*{EndIndent}
\begin{algorithm}[t!]
\caption{\ac{ConMLO}}\label{alg:dcf_11ax}
\begin{algorithmic}[1]
    \State \textbf{Initialize:} $N_\text{tx} = 0$, $\mathcal{L}' = \mathcal{L}$, $\mathcal{L}_w = \emptyset$, $S_{l\in\mathcal{L}}=\text{SENSE}$
    \For{$t = 0, ..., T$}
        \For{$l \in \mathcal{L}'$}
            \Switch{$S_l$}
                \Case{TRANSMIT}
                    \If{$t == (t_0 + T_\text{tx} - \Delta$)}
                        \State \texttt{Start-BO($\mathcal{L}'$)}
                    \EndIf
                    \If{$t == (t_0 + T_\text{tx}$)}
                        \State $N_\text{tx} \leftarrow N_\text{tx} + 1$
                        \State $\mathcal{L}' \cup l$
                        \State $\texttt{Start-BO(}l\texttt{)}$
                    \EndIf
                \EndCase
               \Case{SENSE}    
                    \State Update $P_{int}$
                    \If{$\mathcal{P}_{int} < \phi$} 
                        \State $BO_l \leftarrow BO_l - 1$    
                    \EndIf 
                    \If{$BO_l == 0$}
                        \State $\mathcal{L}_w \cup l$
                    \EndIf
               \EndCase    
               \Case{WAIT}    
                    \State Do nothing.   
               \EndCase   
            \EndSwitch
        \EndFor 
        \If{$\mathcal{L}_w \neq \emptyset$}
            \State $l_w \xleftarrow{R} \mathcal{L}_w$
            \State $\mathcal{L}' \leftarrow \mathcal{L} \setminus l_w$
            \State $S_{l_w} \leftarrow \text{TRANSMIT}$
            \State $S_{l \in \mathcal{L}'} \leftarrow \text{WAIT}$
            \State $\mathcal{L}_w \leftarrow \emptyset$
        \EndIf
    \EndFor
    \Procedure{\texttt{Start-BO($\mathcal{L}'$)}}{}
        \For{$l \in \mathcal{L}'$}
            \State $BO_l\sim U(0, CW)$ 
            \State $S_l \leftarrow \text{SENSE}$
        \EndFor
    \EndProcedure    
\end{algorithmic}
\end{algorithm}

\section{Performance Evaluation}
\label{PerformanceEvaluation}
\thispagestyle{empty} 
\subsection{\ac{WACA} dataset}

\begin{figure}[t!]
    \centering
    \includegraphics[width=.99\linewidth]{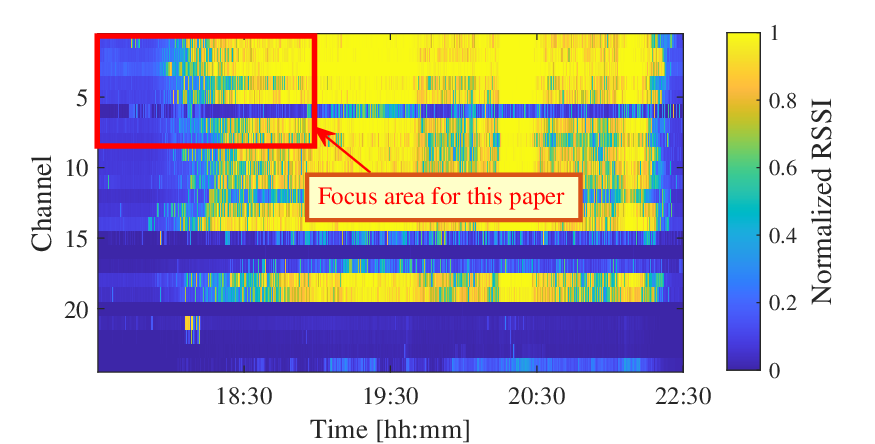}
    \caption{Channel occupancy in \textit{Camp-Nou} (Barcelona, Spain) during a sold-out football match. The red box identifies the portion of the data utilized in this paper.}
    \label{fig:AvCH_OccupancyCampNou}
\end{figure}

The \ac{WACA} dataset~\cite{barrachina2021wi} contains measurements taken at different locations (including university campuses, residential buildings, offices, and a football stadium in a sold-out match) of the full 5~GHz (5170-5815 MHz). The measurements represent the \ac{RSSI} in each scanned 20 MHz channel so that their binary occupancy $\rho\in[0,1]$ (i.e., idle or busy) can be derived using a given \ac{ED} threshold. In this paper, we focus only on a football stadium scenario (namely, \textit{Camp-Nou}), which is the most challenging one in terms of user activity (ranging from idle to fully occupation stages), to assess our \ac{ConMLO} solution. The \textit{Camp-Nou} scenario contains 2000 traces of one second, each of them including 100000 consecutive 10~$\mu$s \ac{RSSI} measurements, in a span of 5 hours. \figref{fig:AvCH_OccupancyCampNou} shows the \textit{Camp-Nou} measurements in terms of the normalized \ac{RSSI}. In our analysis, we focus on the first 750 iterations (of one second each), picking six channels among the first eight. It can be noted that this scenario exhibits a very high a rich diversity in terms of channel occupancy, ranging from empty to saturated channel values. 
Note that we assume that there is no cross-channel interference between transmission and reception between adjacent channels of the \ac{WACA} dataset.

\subsection{Experimental Setup}

In order to evaluate the proposed \ac{ConMLO} scheme, we have simulated the channel access method described in algorithm \ref{alg:dcf_11ax}, and the legacy \ac{SLO} and \ac{MLO} Wi-Fi 7 schemes. 
We have rounded the IEEE 802.11 basic slot duration $\delta$ to 10~$\mu$s, so the duration of channel access parameters such as $T_\text{e}$, SIFS, and DIFS is defined accordingly \cite{barrachina2021wi}. In addition, once the simulated device wins a \ac{TXOP} as a result of the backoff procedure, we assume that such a device can successfully transmit for the maximum time defined by the \ac{TXOP} limit, $T_\text{tx}$. Table~\ref{tab:simulation_parameters} summarizes the main simulation parameters.

\begin{table}[t!]
\centering
\caption{Simulation parameters.}
\label{tab:simulation_parameters}
\begin{tabular}{@{}cccc@{}}
\toprule
\textbf{Parameter} & \textbf{Value} & \textbf{Parameter} & \textbf{Value} \\ \midrule
$T_\text{e}$ & 10~$\mu$s & $|\mathcal{L}|$ & \{1, 2, 4, 6\} \\
DIFS & 30~$\mu$s & $CW$ & 8 \\
SIFS & 10~$\mu$s &  $\phi$ & -82 dBm \\
$T_\text{tx}$ & 5~ms &Traffic model & Full buffer \\
\bottomrule
\end{tabular}
\end{table}

To assess the performance of \ac{ConMLO} and compare it with the baseline \ac{SLO} and Wi-Fi 7 MLO, we focus on:
\begin{itemize}
    \item \textbf{Airtime, $A$:} The percentage of the time the device of interest occupies the channel, computed as ($N_\text{tx}\cdot T_\text{tx})/T$.
    \item \textbf{Consecutive channel acquisitions, $N_\text{ca, cont}$:} The number of consecutive channel acquisitions the device in \ac{ConMLO} was able to perform during the evaluated trace of one second length. 
    %
    \item \textbf{Channel acquisition duration, $T_\text{ca, cont}$:} The duration that \ac{ConMLO} holds a continuous channel access.
\end{itemize}

\subsection{Simulation results}

\thispagestyle{empty} 

\subsubsection{\ac{ConMLO}'s ability to maintain continuity}

\begin{figure}[t!]
    \centering
    \includegraphics[width=.95\columnwidth]   {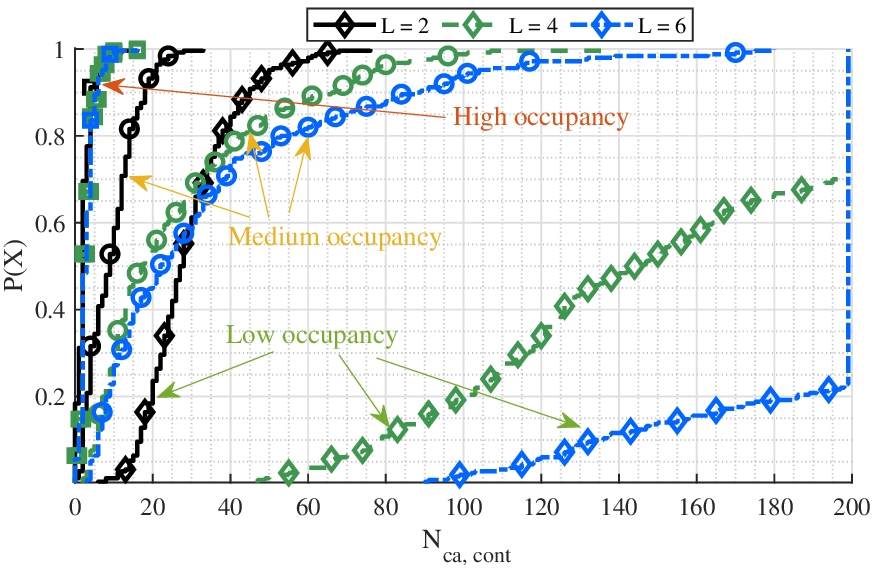}
    \caption{Probability of consecutive $N_{ca, cont}$ for low, medium and high spectrum activity in one second, corresponding to the length of the dataset trace.}
    \label{fig:ContTXwithoutInterruption}
\end{figure}

\begin{figure}[t!]
    \centering
    \includegraphics[width=.95\columnwidth]{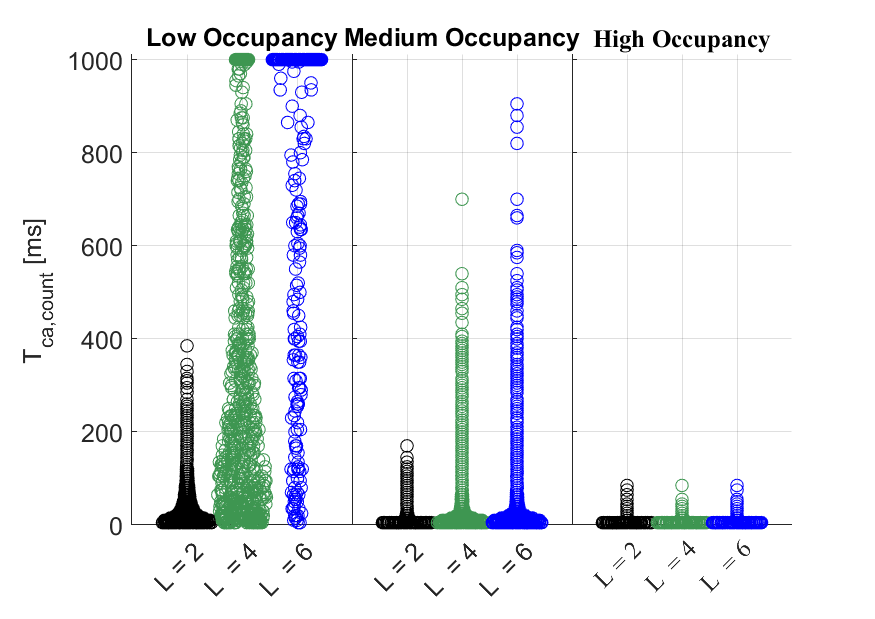}
    \caption{Length distribution of continuous channel accesses experienced by a \ac{ConMLO} in different spectrum activity: low, medium, high.}
    \label{fig:distributionContSTRwithoutInterruption}
\end{figure}

We first consider the performance of a single \ac{ConMLO} device when operating in \textit{Camp-Nou} (see \figref{fig:AvCH_OccupancyCampNou}), from which we distinguish the iterations between low (first 250), medium (251-500) and high occupancy (501-750).
Taking advantage of the degree of freedom introduced by \ac{MLO}, \ac{ConMLO} is able to achieve high values of airtime (above 95\%) for each occupancy, as shown in Table~\ref{tab:airtime_efficiency}.
To analyze the continuity among these transmissions over the medium, \figref{fig:ContTXwithoutInterruption} shows the CDF of $N_\text{ca, cont}$ against the number of available links.
Note that a device reaching $N_\text{ca}^\text{MAX}$=200 has accessed the channel continuously for the entire 1-second trace. 
It can be seen that, for low occupancy and $L=6$, \ac{ConMLO} provides no interruptions almost 80$\%$ of the time. For medium and high occupancy, the difficulty of maintaining continuous access to the channel is accentuated. However, in the worst case with only two links ($L=2$) and high occupancy, \ac{ConMLO} ensures $N_\text{ca, cont} = 4$ in the 30$\%$ of the cases, thus allowing for 20~ms of uninterrupted transmission. The distribution of channel access duration ($T_\text{ca, cont}$) is shown in \figref{fig:distributionContSTRwithoutInterruption} for the different studied cases. As shown, \ac{ConMLO} allows holding the channel for up to 1000~ms (low occupancy, $L=6$), equal to the maximum number of channel acquisitions multiplied by $T_\text{tx}$.
The achievable continuous duration decreases as the environmental channel occupancy increases, thus leading to between 5~ms, one $T_\text{tx}$, and 100~ms values of $T_\text{ca, cont}$.

\begin{table}[t!]
\centering
\caption{Airtime ($A$) achieved by \ac{ConMLO} for different numbers of available channels and spectrum occupancy.}
\label{tab:airtime_efficiency}
\resizebox{.8\columnwidth}{!}{%
\begin{tabular}{@{}l|ccc@{}}
\toprule
 & $|\mathcal{L}| = 2$ & $|\mathcal{L}| = 4$ & $|\mathcal{L}| = 6$ \\ \cmidrule(r){1-4}
Low occupancy & 0.9967 & 0.9999 & 1 \\
Medium occupancy & 0.9803 &  0.9917 & 0.9932  \\
High occupancy & 0.9567 & 0.9610 & 0.9613 \\
\bottomrule
\end{tabular}%
}
\end{table}

\subsubsection{\ac{ConMLO}'s fairness with respect to legacy devices}
\begin{figure}[t!]
    \centering
    \includegraphics[width=0.9\columnwidth]{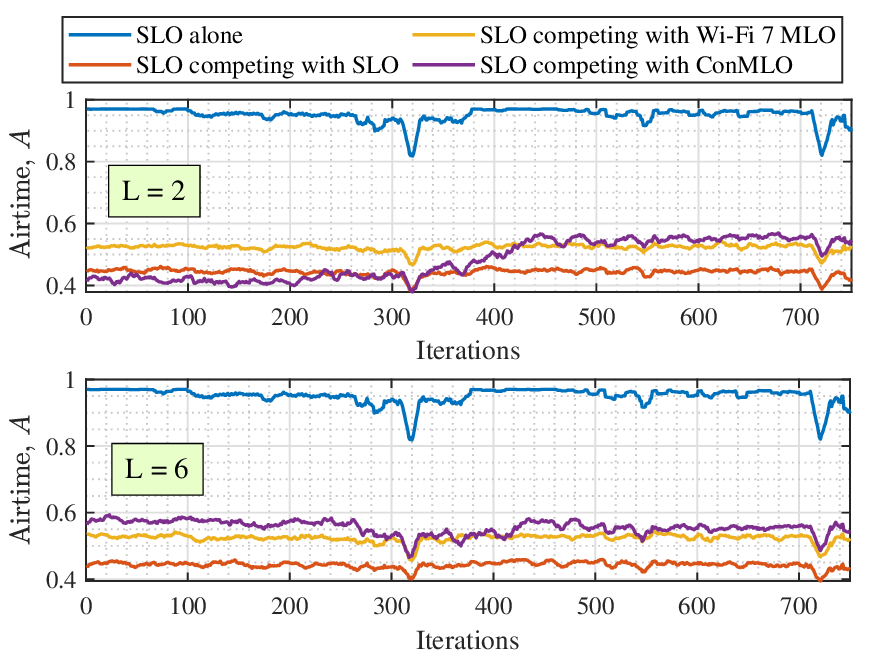}
    \caption{Airtime of a single link device in time versus different competing devices: SLO, Wi-Fi 7 MLO and ConMLO. Top: $L = 2$, bottom: $L = 6$.}  \label{fig:airtimeSingleChannel_vs_othersTXmode}
\end{figure}

\begin{figure}[t!]
    \centering
    \includegraphics[width=0.9\columnwidth]{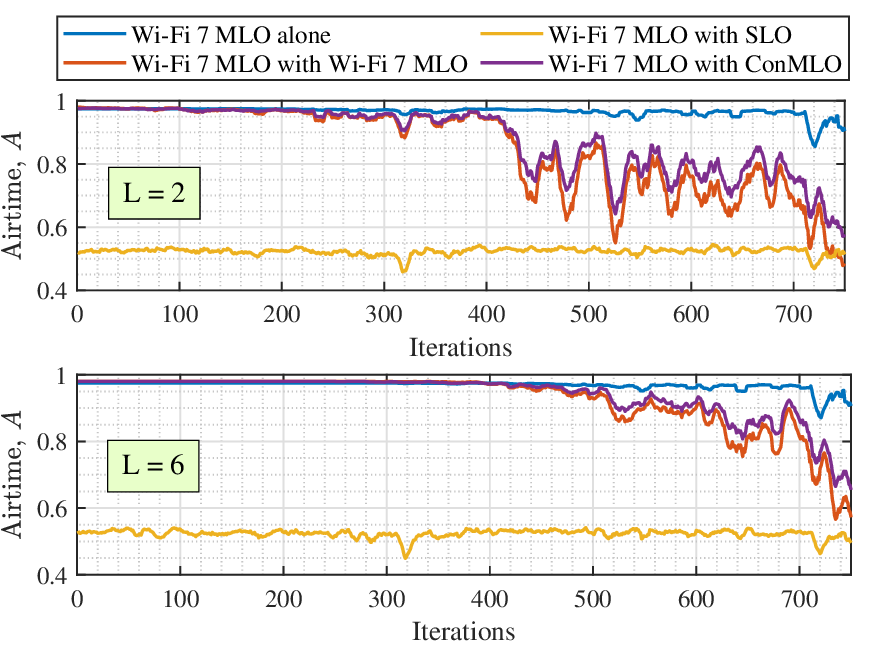}
    \caption{Airtime of a Wi-Fi 7 MLO device in time versus different competing devices: SLO, Wi-Fi 7 MLO and ConMLO. Top: $L = 2$, bottom: $L = 6$.} 
    \label{fig:airtimeMultiLink_vs_othersTXmode}
\end{figure}

Another important aspect of any new method modifying the way wireless systems are operating in the unlicensed spectrum is its fairness when competing for the channel with legacy devices operating in the same band, namely \ac{SLO} and Wi-Fi~7 \ac{MLO} devices.
Note that for this set of results, the interference caused by the competing device is added to the interference measured in the WACA data set.
In particular, as soon as a device is able to access the channel, the WACA dataset is modified to reflect the channel occupancy during the entire $T_\text{tx}$.
For that, we assess the performance of the legacy device in terms of airtime when sharing the channel with other types of devices:

\begin{itemize}
\item \textbf{Impact on \ac{SLO} performance:} \figref{fig:airtimeSingleChannel_vs_othersTXmode} shows the airtime of an \ac{SLO} legacy device in the presence of another \ac{SLO}, a Wi-Fi 7 \ac{MLO}, and a \ac{ConMLO} (being the last two equipped with two and six channels). For completeness, we also show the best-case performance for the legacy \ac{SLO} device (blue straight line), which depicts the case with no competition. As shown, the worst competitor for the \ac{SLO} legacy device is another \ac{SLO} device operating in the same channel. When it comes to multi-link competitors, we find that the \ac{ConMLO} allows the \ac{SLO} device achieving a better performance than when competing with Wi-Fi 7 \ac{MLO} for $L=6$. This result is motivated by the fact that, once the \ac{ConMLO} has acquired a channel for $T_\text{tx}$, it cannot contend on the same channel, thus leaving room for the \ac{SLO} device to operate on its channel.
A rise in the SLO performance is visible at iteration 350 $L=2$, due to decrease in occupancy of the SLO channel.
\item \textbf{Impact on Wi-Fi 7 \ac{MLO} performance:} \figref{fig:airtimeMultiLink_vs_othersTXmode} shows the airtime of a Wi-Fi~7 \ac{MLO} device over time against the same set of competing considered above. In this case, for both $L=2$ and $L=6$ channels, competing for channel access with \ac{ConMLO} (purple line) does not degrade the performance of the Wi-Fi~7 \ac{MLO} device when compared to competition with another Wi-Fi~7 \ac{MLO} device (red line).
\end{itemize}

\section{Conclusions}
\thispagestyle{empty} 
\label{section:conclusions}

In this paper, we introduce and study \ac{ConMLO}, an innovative method to leverage \ac{MLO}, as initially defined in Wi-Fi 7, towards achieving continuous and contention-free channel access in the unlicensed. Our proposed method has been proven its effectiveness at enabling consecutive transmission opportunities in a real scenario. Specifically, our simulations results have shown that \ac{ConMLO} guarantees contention-free transmissions during at least one second in 80\% of the times when six links could be operated in low-loaded scenarios. Furthermore, even for medium to high occupancy conditions, the proposed method is capable of guaranteeing, in average, up to 20 consecutive \ac{TXOP} with 6 channels. In addition, we demonstrated the fairness of \ac{ConMLO} in accessing the spectrum resources with respect legacy \ac{SLO} and Wi-Fi~7 \ac{MLO}. Future works will further investigate the expected performance advantages and limits.

\bibliographystyle{IEEEtran}
\bibliography{biblio}

\end{document}